\documentclass[aps, prl, reprint, superscriptaddress, longbibliography]{revtex4-2}
\usepackage{amsmath}
\usepackage{amssymb}
\usepackage{bm}

\usepackage{graphicx}

\usepackage[version=4]{mhchem}

\usepackage{siunitx}

\usepackage{xcolor}

\begin{document}

%\title{Large-Scale Machine-Learned Molecular Dynamics of Competing Dipolar and Spin-Rotation Relaxation in Solid NH$_4$I}
\title{Angular Momentum Fluctuations Induced by Phonon-Rotation Coupling Govern Nuclear Spin Relaxation in a Molecular Rotator Phase}
%\title{Spin-Rotation Coupling Governs Nuclear Spin Relaxation in Rotator Phase}
%\title{Spin–Rotation and Phonon–Rotation Coupling Resolve Nuclear Spin Relaxation in a Molecular Rotator Phase}

\author{Yanan Li}
\affiliation{Department of Chemistry, New York University,
100 Washington Square East, New York, New York 10003, USA}

\author{Florin Teleanu}
\affiliation{Department of Chemistry, New York University,
100 Washington Square East, New York, New York 10003, USA}
\affiliation{ELI-NP, ``Horia Hulubei'' National Institute for Physics
and Nuclear Engineering, 30 Reactorului Street,
Bucharest-M\u{a}gurele, Ilfov 077125, Romania}

\author{Alexej Jerschow}
\email{alexej.jerschow@nyu.edu}
\affiliation{Department of Chemistry, New York University,
100 Washington Square East, New York, New York 10003, USA}

\date{\today}

\begin{abstract}
Nuclear spin relaxation has been widely used to probe molecular reorientation in solids, yet its interpretation has always required assuming both the relaxation mechanism and the form of the underlying correlation function. In rotator-phase solids more unknowns enter the picture: in ammonium iodide, the proton T$_1$ passes through a maximum and decreases with temperature, a behavior attributed to spin–rotation coupling. Here we model both the dipolar and the spin-rotation contributions to $^1$H relaxation in solid NH$_4$I from first principles, by combining molecular dynamics with machine-learned force fields trained on density functionals and ab initio spin-rotation coupling tensors. The computed rates reproduce the measured values above the order-disorder transition, together with their temperature dependence and frequency independence, and show that fluctuations of the NH$_4^+$ angular momentum, dominates relaxation throughout the rotator phase. The orientational and angular-velocity correlation functions obtained directly from the simulations further share an oscillatory component at lattice-phonon frequencies, revealing a coupling between lattice vibrations and cation reorientation that contributes measurably to relaxation. Separating this component yields a reorientational barrier in agreement with neutron scattering. The approach makes nuclear spin relaxation a quantitative benchmark for simulated dynamics in molecular solids and provides direct access to angular-momentum correlations in the solid state.

% insert abstract here
%Ammonium halides are the textbook system for nuclear spin relaxation in solids; rotational barriers and correlation times for NH$_4^+$ have been extracted from T$_1$ measurements for seventy years using rotational jump models. Above the order–disorder transition, however, no consistent picture exists that correlates polarization lifetimes to molecular dynamics. At these high temperatures, proton relaxation times pass through a maximum that no reorientational model can account for. 
%This has led to conflicting interpretations such as non-thermal barriers or ammonium self-diffusion. 
%Here, we show that in the rotator phase of solid NH$_4$I at temperatures above 270 K, the dominant proton relaxation channel is not $^1$H-$^1$H dipolar coupling modulated by reorientation but spin rotation driven by fluctuations of angular-momentum which carry the opposite temperature dependence. With no adjustable parameters, our forward-prediction workflow based on molecular dynamics using machine-learning force fields quantitatively reproduces the measured relaxation rates on temperatures ranging from 275 to 475 K. These findings highlight the significant contribution of spin-rotation to nuclear spin relaxation in solid-state, motivating the need for a bottom-up predictive approach rather than model fitting. 
\end{abstract}

\keywords{}

\maketitle

Nuclear magnetic resonance (NMR) relaxation arises from time-dependent fluctuations in nuclear spin interactions induced by molecular motion, rendering it highly sensitive to local dynamics~\cite{ernst2023relaxation}. Relaxation theory therefore provides a quantitative bridge between experimentally measured relaxation rates and the underlying microscopic dynamics, enabling the determination of characteristic timescales, activation barriers, and motional mechanisms. For decades, the standard approach has been fitting nuclear polarization lifetimes with analytical equations derived from the Bloembergen-Purcell-Pound (BPP) model\cite{Bloembergen_Purcell_Pound_1948} which assumes that the interactions leading to spin relaxation are governed by rotational jumps. However, recent work\cite{Asthagiri_Chapman_Hirasaki_Singer_2020,Singer_Asthagiri_Chapman_Hirasaki_2017,li2026IL} highlighted the limitations of the BPP model relying on several drastic assumptions that obscure the molecular complexity of chemical systems, supporting the need for a predictive bottom-up workflow to estimate polarization lifetimes from first-principles. 

Ammonium halides have been extensively studied since the early days of NMR due to several factors: i) four equivalent protons on a rigid tetrahedron ii) reorientation rates spanning the NMR window iii) known crystal structures and iv) a classic order-disorder transition. They have served as the benchmark for reorientational relaxation in solids\cite{Itoh_Yamagata_1962,Ogg_Ray_1957,Woessner_Snowden_1967,Sharp_Vrscaj_Pintar_1970}. The rotational barriers have been derived by measuring the proton T$_1$ values at multiple temperatures and fields, assuming the proton relaxation is dominated by $^1$H-$^1$H dipolar (DD) interaction \cite{Mandema_Trappeniers_1974a,Mandema_Trappeniers_1974b,Sharp_Pintar_1976}. However, above the order-disorder phase transition, the T$_1$ values plateau or start decreasing with increasing temperature. Several hypotheses have been advanced such as translational self-diffusion on NH$_4^+$\cite{Sharp_Pintar_1976},  cross-relaxation to quadrupolar nuclei\cite{Sharp_Pintar_1981} and near-free rotation in the rotator phase\cite{Itoh_Yamagata_1962}. Spin rotation (SR) has been proposed as a plausible explanation\cite{sharp1970spin,KOZLENKO_LEWICKI_WĄSICKI_KOZAK_NAWROCIK_SAVENKO_2001}, though quantitative analysis has been lacking. Generally, the role of SR relaxation in solids remains controversial and understudied, as this mechanism is often assumed to be negligible in the solid state.

In this work, we show that SR is the dominant relaxation pathway in ammonium iodide (NH$_4$I) above room-temperature by employing machine-learning force fields (MLFFs) trained on ab-initio molecular dynamics (AIMD) to enable large-scale simulations that accurately reproduce experimentally measured $^1$H relaxation rates. NMR relaxation serves as a quantitative dynamical benchmark for assessing both the fidelity of MLFFs to AIMD and the agreement between simulation and experiment. We systematically examine the influence of exchange-correlation (XC) functionals on hydrogen-bond-mediated NH$_4^+$ dynamics and quantify the DD and SR contributions to relaxation, resolving their relative magnitudes and competition. Vibrational spectral analysis, together with the  dynamical models point to  coupling between phonon and NH$_4^+$ reorientation, which is the origin of oscillations in the SR autocorrelation function (ACF). These results suggest that SR relaxation is significantly more  sensitive to lattice vibrations than the DD interaction. A further supporting fact is that the activation energy for rotation, when corrected for vibration-rotation coupling agrees better with previous neutron scattering factor~\cite{Ren_Qi_Yu_Zhang_Song_Song_Yuan_Wang_Ren_Zhang_et_al._2022}. These results provide guidance for evaluating SR relaxation in solids and clarify the role of phonons in nuclear spin relaxation, a workflow that can be adapted to study more complex materials such as solid-state electrolytes \cite{Pecher_Carretero-González_Griffith_Grey_2017,Zhang_Nazar_2022}.
% In our previous studies~\cite{Flori_pyruvic, li2026IL}, we demonstrated that an MD-based bottom-up framework can connect microscopic motion with experimentally observed relaxation dispersion, highlighting the potential of NMR relaxation as a sensitive benchmark for force-field dynamical validation. Its successful application to complex liquids motivates extending this framework to solid-state systems, where lattice confinement generally restricts translational motion, while molecular reorientation, ionic hopping, and defect migration can remain active. Characterizing these microscopic processes is essential for establishing structure-property relationships and guiding the rational design of functional materials.

\begin{figure}[t]
\centering
\includegraphics[width=\linewidth]{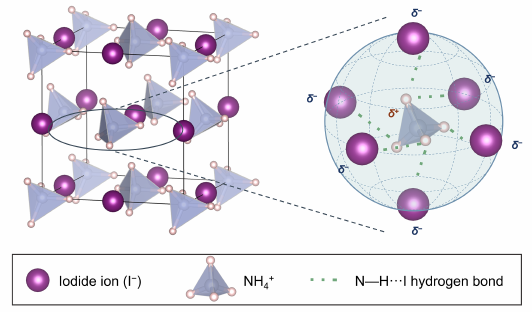}
\caption{
\textbf{Left:} Primitive-cell structure of the $\alpha$ phase of NH$_4$I.
\textbf{Right:} Enlarged view of the local environment of the NH$_4^+$ ion, showing the surrounding I$^-$ ions distributed around the cation and forming an approximately spherical ionic cage.
}\label{fig1}
\end{figure}

Ammonium iodide is a typical rotationally active molecular solid. Above 256 K, it shows a NaCl-type structure with space group $Fm\bar{3}m$~\cite{bartlett1921crystal,Stephenson_Landers_Cole_1952}, in which the relatively low reorientational barrier of NH$_4^+$ gives rise to substantial rotational freedom ("rotator phase"), as illustrated in the left panel of Fig.~\ref{fig1}. Each NH$_4^+$ ion is surrounded by six I$^-$ ions, forming an ionic cage, as shown in the right panel of Fig.~\ref{fig1}. The $\mathrm{N-H\cdots I}$ hydrogen-bonding interactions within this ionic cage constrain the reorientational dynamics of the NH$_4^+$ ions. Therefore, an accurate description of these hydrogen-bonding interactions is essential for reliably simulating NH$_4^+$ rotational dynamics. In DFT, the XC functional provides an approximate treatment of electron exchange and correlation effects and therefore plays a critical role in determining intermolecular interactions and the resulting molecular dynamics~\cite{XC}. In this work, to obtain an accurate description of NH$_4^+$ reorientational dynamics, we evaluated and compared four XC functionals spanning different levels of theory: the GGA functional PBE, the meta-GGA functional r$^2$SCAN, and the hybrid functionals PBE0 and r$^2$SCANh. 

%In the rotator phase of NH$_4$I, $^1$H relaxation rates report an activation energy of 24 meV while $^{127}$I in the same sample reports zero\cite{Sharp_Pintar_1981}. Such inconsistency arose from the usage of the BPP model considering the dipolar and quadrupolar mechanisms modulated by tensor reorientation. 

%Such local reorientation naturally contributes to both $^1$H dipole-dipole (DD) and spin-rotation (SR) relaxation~\cite{hubbard1963theory}, which represent two competing relaxation pathways with distinct dependencies on the molecular reorientational timescale. However, the role of SR relaxation in solids remains controversial, as this mechanism is often assumed to be negligible in the solid state. Although spin relaxation in NH$_4$I has been extensively studied and SR interactions have been proposed to play an important role in spin-lattice relaxation~\cite{sharp1970spin}, the microscopic origin and dynamical pathway of this relaxation mechanism remain incompletely understood. As a simple and well-defined rotational solid, NH$_4$I can serve as a model platform for developing and validating solid-state relaxation theories that can be extended to more complex solid materials, such as battery electrolytes \cite{Pecher_Carretero-González_Griffith_Grey_2017}.

\begin{figure*}[t]
\centering
\includegraphics[width=1\linewidth]{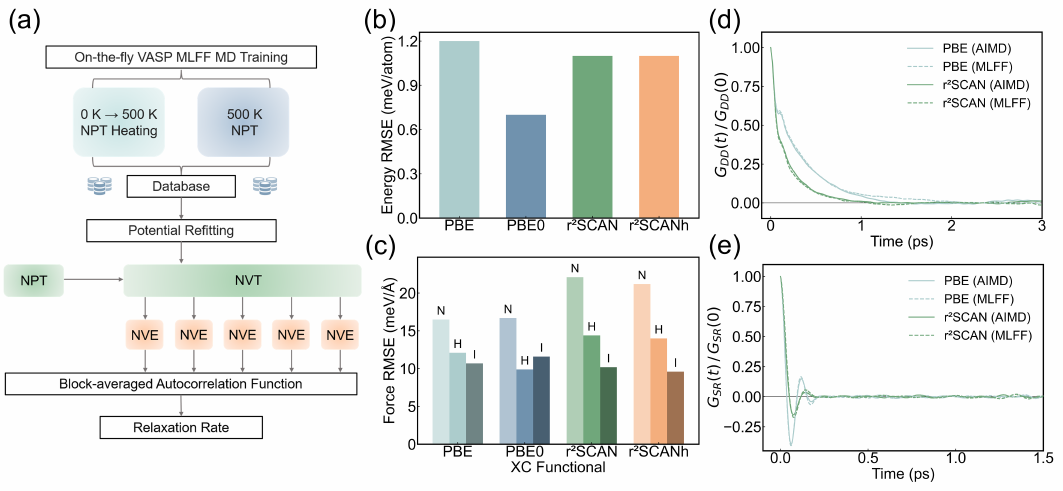}
\caption{
(a) \textbf{Workflow} for training the VASP-MLFF models, refinement of the trained force fields, production molecular dynamics simulations, and subsequent spin-relaxation analysis.
(b,c) \textbf{Energy and force validation:} comparison of the MLFF predictions with DFT calculations, with the corresponding RMSE values reported for the four XC functionals considered in this work.
(d,e) \textbf{Dynamical validation:} comparison of the dipolar and spin-rotation autocorrelation functions obtained from AIMD and MLFF-MD trajectories, respectively.
}\label{fig2}
\end{figure*}

Given the high computational cost of meta-GGA and hybrid functionals, particularly in AIMD simulations, we employed VASP-MLFF to reduce the computational expense and access longer simulation timescales. The VASP-MLFF learns the DFT potential-energy surface from rotationally invariant descriptors of local atomic environments~\cite{VASP_MLFF}. During on-the-fly training, Bayesian force uncertainties are used to identify poorly represented configurations and trigger additional first-principles calculations for iterative refinement of the force field~\cite{MLFF_onthefly, MLFF_onthefly_Bayesian}. Fig.~\ref{fig2}(a) illustrates the overall workflow employed in this study. The trained MLFF was then used for production simulations. Sequential NPT and NVT runs were performed to equilibrate the system. To improve statistical sampling, five initial configurations were extracted from the NVT trajectory and used to initiate independent NVE simulations. The resulting trajectories were then used to calculate block-averaged autocorrelation functions and the following relaxation analysis. To minimize finite size effects, simulations and the corresponding relaxation calculations were additionally performed using $2\times2\times2$, $3\times3\times3$, and $4\times4\times4$ supercells, as summarized in Fig.~S1. Based on this convergence analysis, a $4\times4\times4$ supercell was adopted for the production simulations and subsequent relaxation calculations.

As shown in Fig.~\ref{fig2}(b-c), the resulting VASP-MLFF models reproduce the DFT energies and forces with comparable accuracy, providing a static validation of the learned potential energy surface. However, the relaxation analysis in this work is intrinsically governed by molecular dynamics, agreement in energies and forces alone is not sufficient to establish the reliability of the predicted relaxation behavior. We therefore performed an additional dynamical validation using AIMD simulations under identical conditions and evaluated the corresponding autocorrelation functions using the same analysis procedure. The excellent agreement between the MLFF and AIMD autocorrelation functions demonstrates that the trained models not only reproduce the underlying DFT energetics but also capture the relevant molecular dynamics required for the subsequent relaxation analysis.

\begin{figure}[t]
\centering
\includegraphics[width=\linewidth]{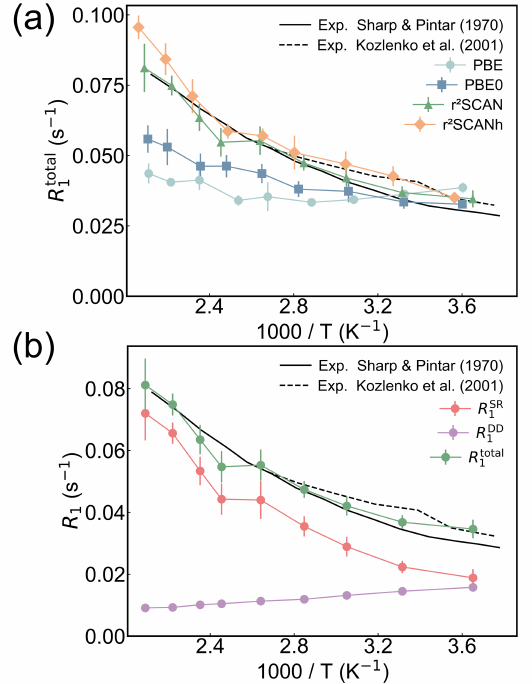}
\caption{
(a) Temperature dependence of the total $^1$H relaxation rate calculated using the four XC functionals and comparison with experimental measurements~\cite{sharp1970spin, KOZLENKO_LEWICKI_WĄSICKI_KOZAK_NAWROCIK_SAVENKO_2001}.
(b) Individual dipolar and spin-rotation contributions to the $^1$H relaxation rate obtained with the r$^2$SCAN functional. Error bars originate from averaging over 5 independent NVE trajectories. 
}\label{fig3}
\end{figure}

As described in supporting information (SI), the reorientational dynamics of NH$_4^+$ give rise to two major spin $^1$H relaxation mechanisms. The first is DD relaxation, which is modulated by fluctuations in the internuclear distances and relative orientations of coupled $^1$H nuclei. The second is SR relaxation, which arises from the coupling between the $^1$H nuclear spins and the molecular rotational angular momentum through the electronic density. As the NH$_4^+$ reorientation becomes faster, the SR contribution becomes increasingly important and can eventually dominate the overall relaxation. A quantitative description of the experimentally observed relaxation therefore requires the DD and SR contributions to be evaluated separately. Such a decomposition also provides direct insight into the underlying molecular dynamics governing the measured nuclear spin relaxation dispersion behavior. For the SR analysis, the N-H bond lengths obtained with different XC functionals were first compared, as shown in Fig.~S2. Among the tested functionals, r$^2$SCAN yields an N-H bond length (1.03~\AA) in closer agreement with the experimental value~\cite{N-H_bond}. The corresponding bond lengths were subsequently used to calculate the molecular moment of inertia required for the SR relaxation analysis, as shown in the Table S1.

The resulting DD, SR, and total relaxation rates obtained from MLFF simulations trained with different XC functionals are presented in Fig.~\ref{fig3}(a). The PBE functional substantially underestimates the relaxation rates, which can be attributed to its prediction of slower NH$_4^+$ reorientational dynamics. As shown in Fig.~S3, PBE exhibits the slowest decay of the DD autocorrelation function, indicating the longest orientational correlation time among the tested functionals. In contrast, the SR autocorrelation function decays most rapidly for PBE, reflecting the distinct dynamical dependence of the SR mechanism on molecular rotation. Among the tested functionals, the meta-GGA r$^2$SCAN not only reproduces the experimental N-H bond length most accurately, but also yields relaxation rates in the closest agreement with experiment. Fig.~\ref{fig3}(b) shows the individual relaxation contributions from SR and DD to the total relaxation rate, revealing that SR relaxation dominates over the investigated temperature range. This result indicates that NH$_4^+$ undergoes remarkably rapid reorientational motion in the crystal, approaching a nearly free-rotor-like dynamical regime despite the surrounding ionic lattice. We further examined two hybrid functionals, PBE0 and r$^2$SCANh, which incorporate 25\% and 10\% Hartree-Fock (HF) exchange, respectively. The PBE0 results fall between those of PBE and r$^2$SCAN, but still underestimate the NH$_4^+$ reorientational dynamics and the corresponding relaxation rates. In contrast, r$^2$SCANh produces slightly faster dynamics and modestly overestimates the relaxation rates relative to experiment.

\begin{figure*}[t]
\centering
\includegraphics[width=1\linewidth]{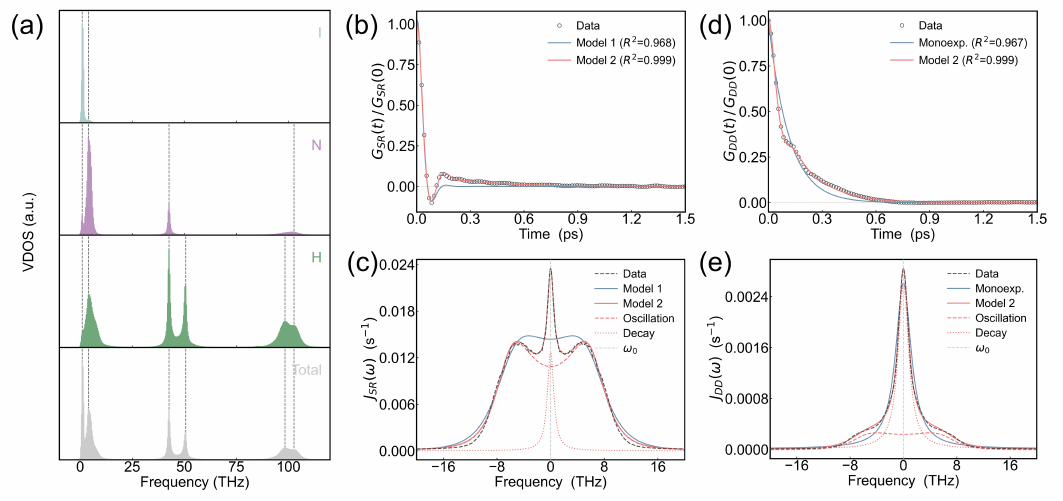}
\caption{
(a) Vibrational density of states (VDOS) of I, N, and H atoms, together with the total VDOS.
(b-e) Dynamical correlations and spectral densities at 300 K: (b) spin-rotation autocorrelation function, (c) spin-rotation spectral density, (d) dipolar autocorrelation function, and (e) dipolar spectral density. Gray circles in (b,d) show the raw autocorrelation function data, while gray dashed curves in (c,e) show their direct numerical Fourier transforms. Solid curves show the model fits. The curves labeled Oscillation and Decay in (c,e) are the Fourier transforms of the Model 2 terms $Ae^{-\Gamma t}\cos(\omega t+\phi)$ and $Be^{-t/\tau_{\mathrm{rot}}}$, respectively.
}\label{fig4}
\end{figure*}

Despite the excellent agreement between the simulated and experimental relaxation rates, several physical features remain to be explained. The SR ACF exhibits obvious oscillations, while the DD ACF displays a subtle "shoulder", both appearing on comparable timescales (Fig.~2d-e). Furthermore, the spectral density obtained by Fourier transform (FT) deviates from a single Lorentzian-shape peak, exhibiting shoulders on either side of the central peak. An Arrhenius analysis of the correlation times (Fig. S4) obtained by integrating the normalized DD ACF also yields an apparent activation energy slightly higher than the experimental value of 25~meV~\cite{Ren_Qi_Yu_Zhang_Song_Song_Yuan_Wang_Ren_Zhang_et_al._2022}. Therefore, understanding these physical phenomena requires deeper insight into NH$_4^+$ reorientational dynamics.

The negative lobes in the SR ACF indicate anticorrelated angular motion, suggesting that NH$_4^+$ reorientation is accompanied by oscillatory motion. To investigate its vibrational origin, we calculated the vibrational density of states (VDOS), as shown in Fig.~\ref{fig4}(a). The peaks at 1.007 and 3.998~THz are assigned to lattice vibrations~\cite{goel1977lattice, vagelatos1975lattice}, whereas those at 42.694 and 50.960~THz correspond to symmetric and asymmetric NH$_4^+$ vibrational modes~\cite{huang2015structural}. The higher-frequency peaks at 98.358 and 103.027~THz are associated with symmetric and asymmetric N-H bond stretching~\cite{huang2015structural}. The large mass of iodine limits its response to NH$_4^+$ motion, resulting in an iodine VDOS dominated by two relatively sharp lattice-vibration peaks. Interestingly, in addition to peaks associated with internal NH$_4^+$ motion, the hydrogen VDOS exhibits noticeably broader lattice-vibration peaks than those in the iodine VDOS. This phonon broadening may reflect coupling between lattice phonons and NH$_4^+$ reorientational motion~\cite{Ren_Qi_Yu_Zhang_Song_Song_Yuan_Wang_Ren_Zhang_et_al._2022, lynden1994translation}.

Motivated by these observations, we propose two possible dynamical models to describe the oscillatory behavior of the SR ACF. Model~1 consists of a damped oscillatory term representing vibrational motion,

\begin{equation}
C_{\mathrm{Model\,1}}(t)
=
A e^{-\Gamma t}\cos(\omega t+\phi),
\label{eq:acf_model1}
\end{equation}
whereas Model~2 includes an additional monoexponential term representing rotational decorrelation,
\begin{equation}
C_{\mathrm{Model\,2}}(t)
=
A e^{-\Gamma t}\cos(\omega t+\phi)
+
B e^{-t/\tau_{\mathrm{rot}}}.
\label{eq:acf_model2}
\end{equation}
Here, $A$ and $B$ are the amplitudes of the lattice-vibrational and rotational components, respectively; $\Gamma$ is the phonon damping rate, $\omega$ is the oscillation angular frequency, $\phi$ is the phase, and $\tau_{\mathrm{rot}}$ is the rotational correlation time. Model~2 therefore provides a phenomenological description of NH$_4^+$ dynamics that includes both lattice vibrations and rotational decorrelation.

As shown in Fig.~\ref{fig4}(b), Model~2 provides a better fit to the SR ACF. Its analytical FT also closely reproduces the spectral density obtained by direct numerical Fourier transformation. Within this model, the rotational component produces the central Lorentzian peak, while the lattice-vibrational component gives rise to the symmetric peaks at around 6~THz. Both components exhibit nonzero spectral density at the Larmor frequency 30 MHz and therefore contribute to the relaxation rate. Notably, the fitted oscillation frequency, $\omega/(2\pi)$, lies within the lattice-vibration peak in the VDOS, supporting the involvement of lattice vibrations in the SR dynamics.

We applied a similar analysis to the DD ACF, comparing Model~2 with a monoexponential model, as expected for isotropic rotational diffusion in the BPP theory. Model~2 again provides a better fit and accurately captures the subtle transient plateau in the DD ACF. The corresponding analytical spectral density closely agrees with that obtained by numerical Fourier transformation of the simulated ACF. Moreover, the similar oscillation frequencies extracted independently from the DD and SR ACFs support a common vibrational contribution to NH$_4^+$ reorientational dynamics, consistent with phonon-reorientation coupling. The more pronounced oscillations in the SR ACF suggest that its sensitivity to vibrational motion makes it a possible probe of vibrational dynamics.

Figure~S7 presents the temperature-dependent contributions of the rotational and vibrational motions to the DD and SR relaxation rates, as resolved using Model~2. Although the rotational component dominates both relaxation mechanisms, the vibrational contribution remains non-negligible, particularly for SR relaxation. Accounting for this contribution is therefore necessary for an accurate description of the calculated relaxation rates.

Finally, we extracted the rotational correlation times from the Model 2, $\tau_{\mathrm{rot}}$, and analyzed their temperature dependence using an Arrhenius fit (Fig. S8). The resulting activation energy, $23.4 \pm 2.3\,\mathrm{meV}$, agrees well with the experimental value of 25~meV derived from quasi-elastic neutron scattering\cite{Ren_Qi_Yu_Zhang_Song_Song_Yuan_Wang_Ren_Zhang_et_al._2022}. This agreement suggests that the slightly higher apparent activation energy obtained by integrating the full normalized DD ACF reflects the combined contributions of rotational and vibrational dynamics. Separating these components therefore enables a more direct comparison of the simulated reorientational barrier with experiment.

In summary, we provide a bottom-up computational approach to solve a decades-old unresolved feature of proton T$_1$ values in the molecular rotator ammonium iodide. 
%We propose NMR relaxation as a sensitive benchmark for assessing the dynamical accuracy of force fields, including machine-learning force fields. Beyond characterizing timescales, relaxation analysis can help distinguish different types of molecular motion.
By combining MLFFs trained using different XC functionals with MD simulations and subsequent relaxation analysis, we find that r$^2$SCAN accurately reproduces both structural properties, such as bond lengths, and NH$_4^+$ reorientational dynamics, consistent with its reliable description of hydrogen-bonding interactions. We find that spin rotation originating from modulation of angular momentum is the dominant relaxation pathway in the rotator phase of NH$_4$I at high temperatures. Further analysis of the vibrational density of states, together with the proposed dynamical models, reveals  coupling between lattice phonons and NH$_4^+$ reorientation. When taken into account for rotational analysis,  the computed activation energy moves into further alignment with experiment. Such studies would also be of relevance in the investigation of the paddlewheel effect~\cite{Zhang_Nazar_2022}. Lastly, we anticipate that the similar relaxation-based validation into MLFF development could help establish models that reliably reproduce both equilibrium properties and experimentally relevant molecular dynamics for more complex systems such as solid-state electrolytes. 

\section{Acknowledgements} 
F.T. acknowledges funding from National Medical Project MySMIS 326475 and Project ELI-RO/RDI/2024/14 SPARC funded by the Institute of Atomic Physics (Romania). A.J. acknowledges funding from the US National Science Foundation, Award No. CHE 2505792, an award from the ACS PRF 68117-ND6, and an unrestricted gift from Google. We acknowledge support from New York University's High Performance Computing resources and staff (especially Dr. Shenglong Wang).  The authors acknowledge stimulating discussions with Prof. Kong Ooi Tan about anomalous dynamics in ammonium halides.

\section{Data Availability Statement}
The processed data supporting this study will be made openly available on Zenodo at https://doi.org/10.5281/zenodo.23107149 upon publication of the article.

%This work highlights the potential of NMR relaxation as a complementary benchmark for developing and validating MLFFs. While MLFFs are typically trained and evaluated against reference energies and forces, relaxation provides an experimentally accessible test of whether this accuracy translates into the molecular dynamics. Combining multiple relaxation mechanisms may help identify deficiencies in the types and timescales of motion reproduced by an MLFF, thereby informing the choice of reference electronic-structure methods and the refinement of training datasets. Future work could extend this approach to complex liquid and solid systems to evaluate the dynamical accuracy and transferability of MLFFs across different thermodynamic conditions and local environments. We believe that the relaxation-based validation into MLFF development could help establish models that reliably reproduce both equilibrium properties and experimentally relevant molecular dynamics.

% If in two-column mode, this environment will change to single-column
% format so that long equations can be displayed. Use
% sparingly.
%\begin{widetext}
% put long equation here
%\end{widetext}

%\appendix
%\section{}

\begin{acknowledgments}

\end{acknowledgments}

\bibliography{Reference.bib}

\end{document}

\setcounter{figure}{0}
\setcounter{table}{0}
\setcounter{equation}{0}

\begin{center}

\vspace*{0.8cm}

{\LARGE \textbf{Supporting Information:}\par}

\vspace{1.2cm}
{\LARGE \textbf{Angular Momentum Fluctuations Induced by Phonon-Rotation Coupling Govern Nuclear Spin Relaxation in a Molecular Rotator Phase}\par}
%{\LARGE \textbf{Large-Scale Machine-Learned Molecular Dynamics of Competing Dipolar and Spin-Rotation Relaxation in Solid NH$_4$I}\par}

\vspace{1.4cm}

{\large
Yanan Li$^{\dagger}$,
Florin Teleanu$^{\dagger,\ddagger}$, 
and Alexej Jerschow$^{*,\dagger}$\par}

\vspace{0.9cm}

{\itshape
$^\dagger$Department of Chemistry, New York University, New York, New York 10003, United States\par}

\vspace{0.3cm}

{\itshape
$^\ddagger$ELI-NP, “Horia Hulubei” National Institute for Physics and Nuclear Engineering, 30 Reactorului Street, Bucharest-Magurele 077125, Ilfov, Romania\par}

\vspace{0.8cm}

E-mail: alexej.jerschow@nyu.edu

\end{center}

\vspace{1cm}

\noindent\textbf{Contents of the Supporting Information}

\vspace{0.3cm}

\noindent This Supporting Information contains:
\begin{itemize}
    \item Theoretical and Computational Methods.
    \item Relaxation theory.
    \item Supercell size convergence tests for SR and DD autocorrelation functions.
    \item Structural parameters and spin-rotation coupling coefficients calculated using different XC functionals.
    \item Comparison of SR and DD autocorrelation functions, correlation times, and activation energies obtained using different XC functionals.
    \item Temperature-dependent SR and DD autocorrelation functions and spectral density functions calculated using r$^2$SCAN.
    \item Ensemble Dependence of Spin-Rotation and Dipolar Relaxation.
    \item Vibrational and rotational contributions to SR and DD relaxation rates determined using Model~2.
    \item Activation energy barrier of rotational motion.
\end{itemize}

\vspace{0.5cm}

\section{Theoretical and Computational Methods}

All molecular simulations were performed using the Vienna \textit{Ab initio} Simulation Package (VASP)~\cite{vasp}, version 6.5.1. The spin-rotation coupling tensors were calculated separately using ORCA~\cite{neese2020orca}.

\subsection{Density Functional Theory Calculations}

The primitive crystal structure was first optimized using DFT. The core-valence electron interactions were described using the projector augmented-wave (PAW) method~\cite{paw}, and a plane-wave energy cutoff of 600 eV was employed. The convergence criteria for the electronic energy and atomic forces were set to $10^{-8}$ eV and 0.01 eV~\AA$^{-1}$, respectively. To assess the sensitivity of the hydrogen-bonding description to the choice of XC approximation, structural optimizations were performed using the PBE generalized gradient approximation (GGA) functional~\cite{GGA}, the r$^2$SCAN meta-GGA functional~\cite{r2scan}, and the hybrid functionals PBE0~\cite{PBE0, PBE0_2} and r$^2$SCANh~\cite{r2scanh}. Brillouin-zone integrations were performed using a $\Gamma$-centered $k$-point mesh with a reciprocal-space spacing of approximately 0.02~\AA$^{-1}$.

For subsequent validation calculations, DFT single-point calculations were performed using a $2\times2\times2$ supercell with a plane-wave energy cutoff of 500 eV. The electronic energy convergence criterion was set to $10^{-5}$ eV.

\subsection{Machine-Learned Force Field Molecular Dynamics}

A separate MLFF was independently trained for each XC functional considered in this work using the VASP-MLFF framework~\cite{VASP_MLFF, MLFF_onthefly, MLFF_onthefly_Bayesian}. Representative configurations for MLFF training were collected using the on-the-fly learning scheme from an NpT MD trajectory. The system was first heated from 0 to 500 K over 10 ps and was then equilibrated at 500 K for an additional 10 ps. Both simulations were performed at zero external pressure with a time step of 1 fs. The NpT simulations were performed using a Langevin thermostat and barostat. The friction coefficients for the atomic degrees of freedom were set to 5 ps$^{-1}$ for N, I, and H atoms, while a friction coefficient of 1 ps$^{-1}$ was applied to the lattice degrees of freedom.

The trained force field was subsequently used in production mode to perform NpT, NVT, and NVE MLFF-MD simulations using a $4\times4\times4$ supercell at fixed temperatures ranging from 275 to 475 K. At each temperature, an initial 50 ps NpT ensemble was performed to relax the simulation box, followed by a 50 ps NVT trajectory for thermal equilibration. Five configurations were then uniformly sampled from the NVT trajectory and used as independent initial configurations for NVE simulations. Five independent NVE trajectories of 4 ps each were generated for subsequent spin relaxation analyses. A time step of 1 fs was used for all molecular dynamics simulations. Finite-size convergence tests were performed using $2\times2\times2$, $3\times3\times3$, and $4\times4\times4$ supercells to minimize possible finite-size effects.

For comparison and validation, conventional AIMD simulations were also performed under the same simulation conditions using a $2\times2\times2$ supercell. The AIMD trajectories were directly compared with MLFF-MD simulations performed using the same supercell size to assess the consistency between the first-principles and MLFF descriptions of the molecular dynamics.

\subsection{Spin-Rotation Coupling Tensor Calculations}

The $\mathrm{NH_4^+}$ geometry used for the spin-rotation coupling calculations was extracted from the optimized primitive cell.Spin-rotation coupling tensors were calculated independently for each XC functional considered in this work using ORCA 6.1.0. The calculations employed the def2-TZVP basis set. Spin-rotation coupling constants were evaluated using the SROT keyword implemented in the EPR/NMR module of ORCA~\cite{ORCA_SR}, as shown in Table S2.

\begin{table}[htbp]
    \centering
    \caption{Molecular moments of inertia of NH$_4^+$ obtained
    using different exchange-correlation functionals.}
    \label{tab:nh4_inertia}
    \begin{tabular}{lc}
        \hline
        Functional
        & $I_{\mathrm{mol}}$ ($10^{-47}\,\mathrm{kg\,m^2}$) \\
        \hline
        PBE        & 4.830629 \\
        r$^2$SCAN  & 4.734536 \\
        PBE0       & 4.719838 \\
        r$^2$SCANh & 4.706996 \\
        \hline
    \end{tabular}
\end{table}

\begin{table}[htbp]
    \centering
    \caption{Spin-rotation coupling constants of NH$_4^+$,
    averaged over the four hydrogen nuclei for each
    exchange-correlation functional. All values are in kHz.}
    \label{tab:nh4_sr_constants}
    \begin{tabular}{lccc}
        \toprule
        Functional
        & $C_{\parallel}$
        & $C_{\perp,1}$
        & $C_{\perp,2}$ \\
        \midrule
        PBE       & 1.0030 & 21.9462 & 22.0020 \\
        r$^2$SCAN & 1.3600 & 22.6453 & 22.7082 \\
        PBE0      & 1.2917 & 22.7877 & 22.8500 \\
        r$^2$SCANh & 1.4303 & 22.8710 & 22.9250 \\
        \bottomrule
    \end{tabular}
\end{table}

\section{Spin Relaxation Theory}

Spin relaxation is widely described within the framework of Redfield theory~\cite{wangsness1953dynamical, redfield1955nuclear, redfield1957theory}, which provides a second-order perturbative treatment of relaxation induced by time-dependent fluctuations of the spin Hamiltonian. Starting from the Liouville-von Neumann equation, the Redfield formalism leads to a relaxation superoperator that describes the evolution of the spin density operator. In a general form, the relaxation superoperator can be written as
\begin{equation}
\hat{\hat{\Gamma}}\{Q\}
=
\frac{1}{2}
\sum_{\mu,\mu'}
\sum_{p,p'}
\left[
V_{p}^{(\mu)},
\left[
\left(V_{p'}^{(\mu')}\right)^{\dagger},
Q
\right]
\right]
J_{p,p'}^{(\mu,\mu')}
\left(\omega_p\right)
\delta\left(\omega_p,\omega_{p'}\right).
\end{equation}
where the double commutator determines the allowed relaxation pathways 
through the corresponding spin operators, while the spectral-density 
function $J_{\ell,\ell,-m,-m}^{\mu,\mu'}(\omega_{p'})$ determines the 
strength of each relaxation channel at the relevant transition frequency~\cite{ernst2023relaxation}. 
This formalism therefore provides a general framework for evaluating NMR 
spin relaxation from the microscopic fluctuations of spin interactions. Since these fluctuations arise from the time-dependent modulation of the spin Hamiltonian by molecular motion, the resulting relaxation rates encode information about the amplitudes and timescales of the underlying dynamics. NMR relaxation can therefore be used as a quantitative probe of molecular motion through the corresponding correlation functions and spectral densities.

\subsection{Spin-Rotation Relaxation}

The SR interaction is described by the following Hamiltonian,
\begin{equation}
\hat{H}_{\mathrm{SR}}
=
\hat{\mathbf{I}}\cdot\mathbf{C}\cdot\hat{\mathbf{J}},
\end{equation}
where $\hat{\mathbf{I}}$ is the nuclear spin angular momentum, 
$\hat{\mathbf{J}}$ is the molecular rotational angular momentum, and 
$\mathbf{C}$ is the SR coupling tensor.  The SR interaction therefore arises from the coupling between 
the nuclear spin and the rotational angular momentum of the molecule, mediated by the SR coupling tensor $\mathbf{C}$. Following the Redfield formalism, the ACF of SR relaxation can be expressed as
\begin{equation}
G_{SR}(t)
=
\frac{I k_{\mathrm{B}}T}{\hbar^2}
\left(
C_a^2+\frac{2}{9}C_d^2
\right)
\left\langle
\omega(t_0)\omega(t_0+t)
\right\rangle_{t_0},
\end{equation}
where I is the molecular moment of inertia, $k_{\mathrm{B}}$ is the Boltzmann constant, and $T$ is the temperature~\cite{singer2018nmr}. The isotropic and anisotropic components of the spin-rotation coupling tensor are defined as
\begin{equation}
C_a=\frac{2C_{\perp}+C_{\parallel}}{3},
\qquad
C_d=C_{\perp}-C_{\parallel},
\end{equation}
where $C_{\perp}$ and $C_{\parallel}$ denote the two principal components of the spin-rotation coupling tensor.

The instantaneous molecular angular velocity $\boldsymbol{\omega}$ was obtained from the atomic positions and site velocities according to
\begin{equation}
\boldsymbol{\omega}\times\mathbf{r}_i=\mathbf{v}_i,
\end{equation}
where $\mathbf{r}_i$ and $\mathbf{v}_i$ are the position and velocity of atom $i$, respectively, relative to the molecular center of mass. Then, $\boldsymbol{\omega}$ was determined by minimizing the least-squares objective function
\begin{equation}
\mathcal{L}
=
\sum_i
\left|
\mathbf{v}_i-\boldsymbol{\omega}\times\mathbf{r}_i
\right|^2.
\end{equation}

The corresponding spectral density function was obtained by applying a Fourier transform (FT) to the autocorrelation function,
\begin{equation}
J_{SR}(\omega)
=
2\int_{0}^{\infty}
G_{SR}(t)\cos(\omega t)\,dt,
\end{equation}
and the spin-rotation contribution to the longitudinal relaxation rate was then evaluated as
\begin{equation}
R_1^{SR}
=
J_{SR}(\omega_0),
\end{equation}
where $\omega_0$ is the nuclear Larmor frequency.

\subsection{Dipolar Relaxation}

The dipolar interaction is described by the following Hamiltonian,
\begin{equation}
\hat{H}_{\mathrm{DD}}
=
\hat{\mathbf{I}}\cdot\mathbf{D}\cdot\hat{\mathbf{S}},
\end{equation}
where $\hat{\mathbf{I}}$ and $\hat{\mathbf{S}}$ are the nuclear spin angular momentum operators of two interacting nuclei, and $\mathbf{D}$ is the dipolar coupling tensor. The dipolar interaction couples neighboring nuclear spins through the dipolar tensor and depends strongly on both the internuclear distance and the orientation of the internuclear vector. Accordingly, the ACF of the DD interaction can be expressed as
\begin{equation}
G_{\mathrm{DD}}(t)
=
\frac{1}{5}
\left(
\frac{\mu_0 \hbar \gamma_I \gamma_S}{4\pi}
\right)^2
S(S+1)
\left\langle
\sum_{P_{IS}}
\frac{
P_2\!\left[
\hat{\mathbf{u}}_{IS}(t_0)\cdot
\hat{\mathbf{u}}_{IS}(t_0+t)
\right]
}{
r_{IS}^3(t_0)\,
r_{IS}^3(t_0+t)
}
\right\rangle,
\label{ACF_DD}
\end{equation}
where $\mu_0$ is the magnetic permeability of vacuum, $\hbar$ is the reduced Planck constant, $\gamma_I$ and $\gamma_S$ are the gyromagnetic ratios of the two spins, $P_{IS}$ denotes the set of \{$I,S$\} spin pairs, $P_2$ is the second-order Legendre polynomial, $\hat{\mathbf{u}}_{IS}$ is the unit vector connecting spins $I$ and $S$, and $r_{IS}$ is the corresponding time-dependent internuclear distance obtained from the MD trajectories~\cite{li2026IL}.

The corresponding spectral density function was obtained by Fourier transforming the autocorrelation function,
\begin{equation}
J_{\mathrm{DD}}(\omega)
=
2\int_0^\infty
G_{\mathrm{DD}}(t)\cos(\omega t)\,dt.
\end{equation}
For the homonuclear case, the longitudinal relaxation rate is given by
\begin{equation}
R_1^{DD}
=
J_{\mathrm{DD}}(\omega_0)
+
4J_{\mathrm{DD}}(2\omega_0).
\end{equation}

In case of NH$_4$I, $^{1}$H dipolar and spin-rotation relaxation associated with the NH$_4^+$ ions exhibit competing dependences on the reorientational dynamics. Rapid reorientation shortens the orientational correlation time and suppresses the dipolar contribution through motional averaging. In contrast, faster rotation increases the magnitude of the molecular angular velocity, while the angular velocity fluctuations may remain correlated over a comparatively longer timescale, thereby enhancing the spin-rotation relaxation contribution. By separately evaluating the dipolar and spin-rotation contributions from the MD trajectories, the total $^{1}$H relaxation rate can be reconstructed and directly compared with experiment. Reproducing the experimental relaxation behavior therefore places stringent constraints on the simulated NH$_4^+$ reorientational dynamics, making it a demanding benchmark for the underlying molecular motion.
\newpage
\section{Supercell Size Convergence}

\begin{figure}[!htbp]
\centering
\includegraphics[width=\linewidth]{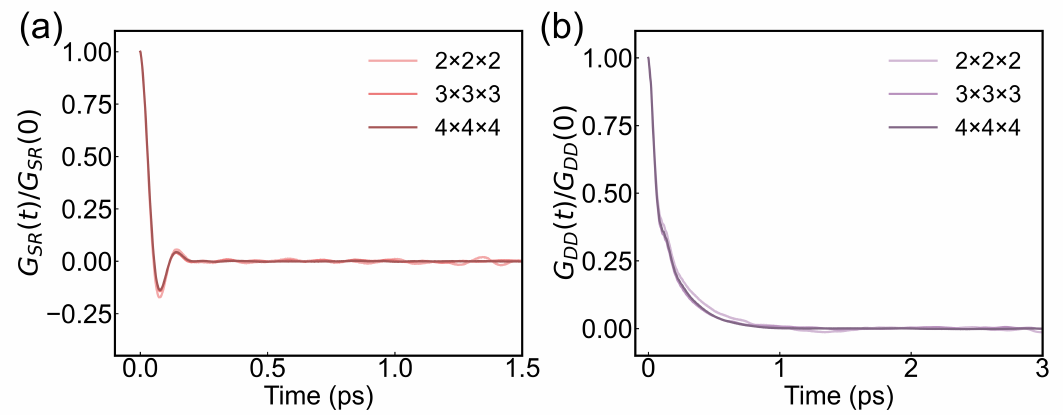}
\caption{(a) SR and (b) DD autocorrelation functions calculated using $2\times2\times2$, $3\times3\times3$, and $4\times4\times4$ supercells.}
\label{figS1}
\end{figure}

\section{Structural Parameters}

\begin{figure}[!htbp]
\centering
\includegraphics[width=\linewidth]{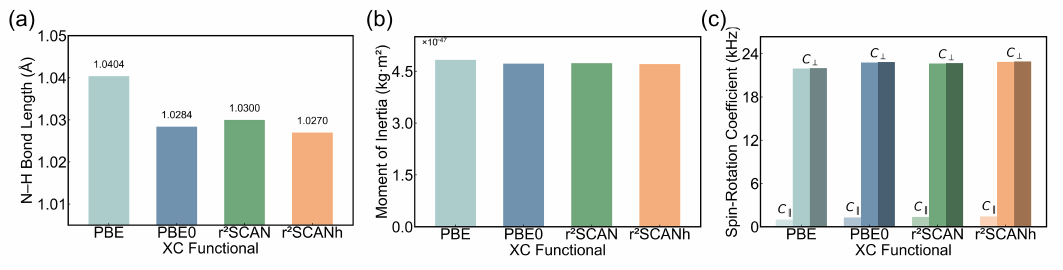}
\caption{(a) Average N-H bond length, (b) molecular moment of inertia, and (c) spin-rotation coupling coefficient of NH$_4^+$ calculated using different XC functionals.}
\label{figS2}
\end{figure}

\newpage
\section{Comparison of Exchange-Correlation Functionals}

\begin{figure}[!htbp]
\centering
\includegraphics[width=\linewidth]{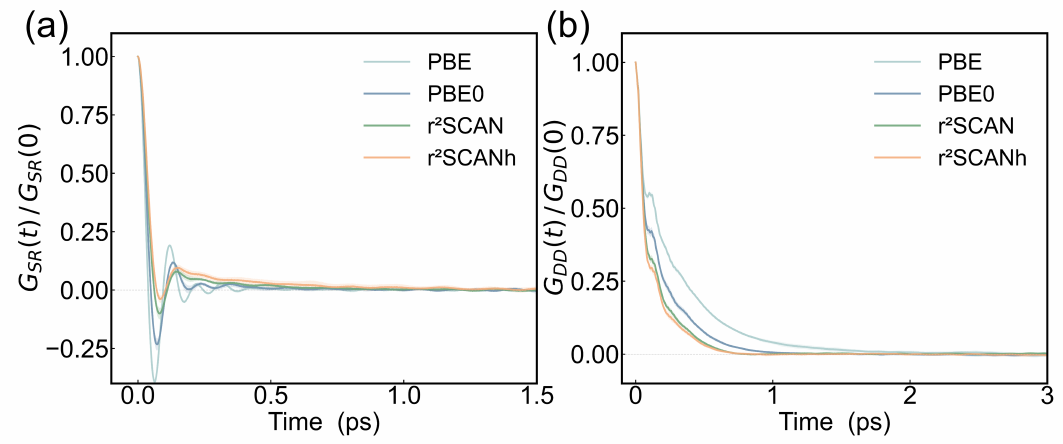}
\caption{(a) SR and (b) DD autocorrelation functions calculated using different XC functionals at 300~K.}
\label{figS3}
\end{figure}

\begin{figure}[!htbp]
\centering
\includegraphics[width=\linewidth]{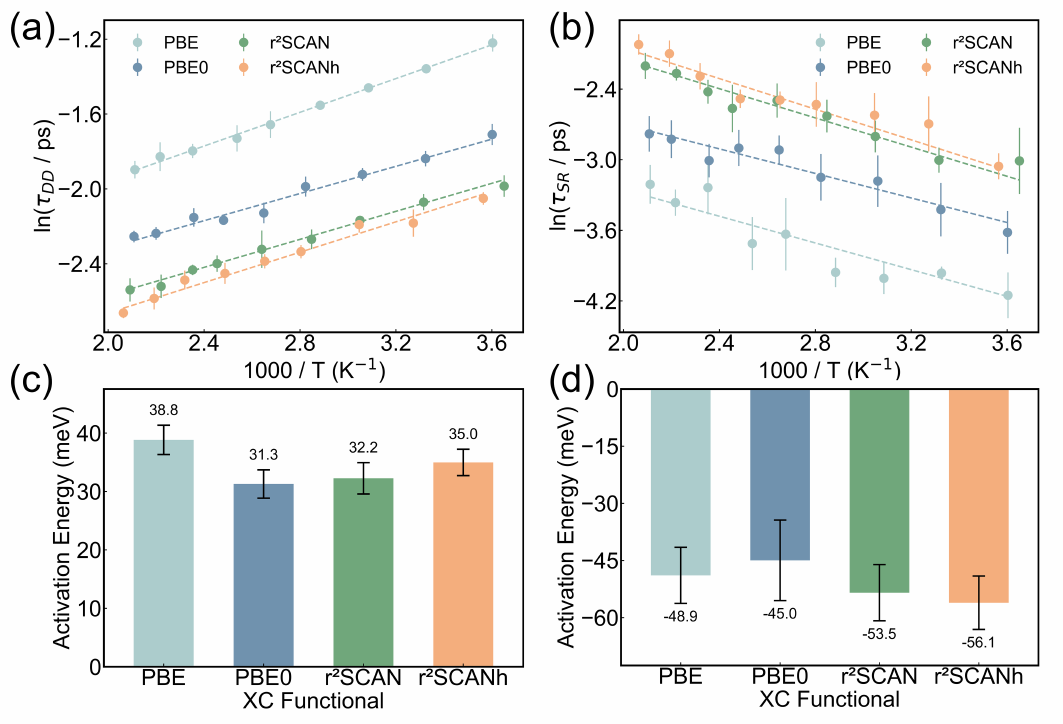}
\caption{Arrhenius plots of (a) DD and (b) SR correlation times calculated using different XC functionals, with the corresponding activation energies shown in (c) and (d), respectively.}
\label{figS4}
\end{figure}

\section{Temperature-Dependent Autocorrelation Functions and Spectral Densities from r$^2$SCAN}

\begin{figure}[!htbp]
\centering
\includegraphics[width=\linewidth]{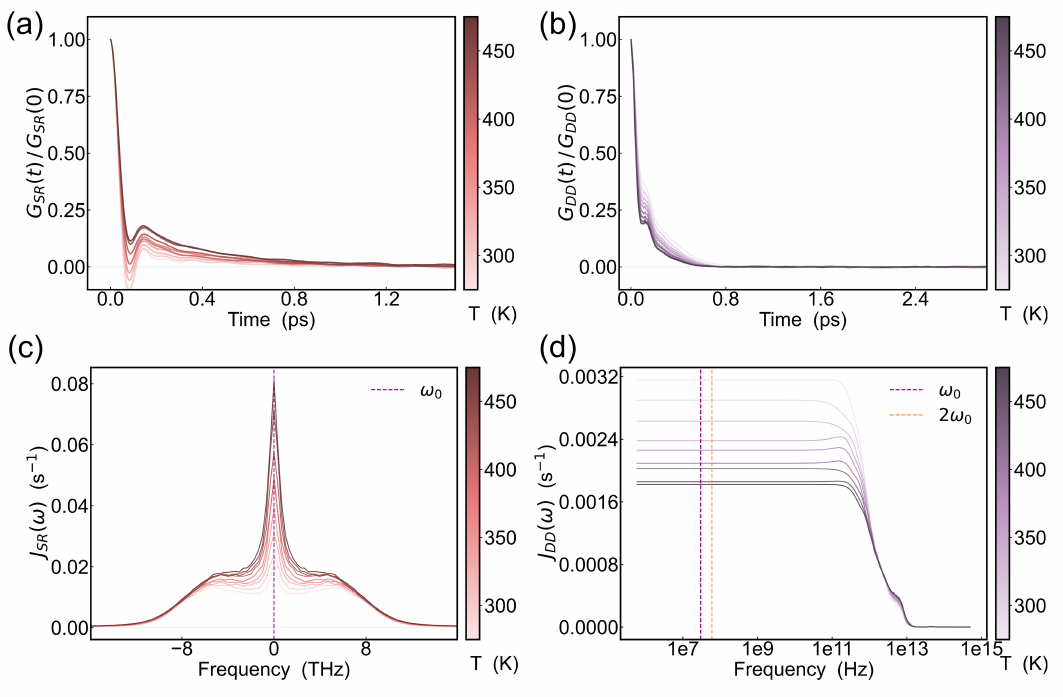}
\caption{(a) SR and (b) DD autocorrelation functions calculated at different temperatures using r$^2$SCAN, together with the corresponding spectral density functions obtained by direct numerical Fourier transformation.}
\label{figS5}
\end{figure}

\newpage
\section{Ensemble Dependence of Spin-Rotation and Dipolar Relaxation}
Fig. S6 compares the DD, SR, and total relaxation rates obtained from NPT, NVT, and NVE simulations using the same XC functional, r$^2$SCAN. The DD contribution remains nearly invariant across the three ensembles, indicating that the orientational fluctuations governing DD relaxation are comparatively insensitive to the choice of ensemble. In contrast, the SR contribution exhibits a substantial ensemble dependence. Although the NPT and NVT simulations yield similar SR relaxation rates, the corresponding NVE values are systematically larger and show closer agreement with experiment. This behavior can be rationalized by the microscopic origin of SR relaxation. The SR correlation function is determined by fluctuations in the molecular angular velocity and is therefore directly sensitive to the instantaneous atomic velocities. Thermostats and barostats introduce additional dynamical perturbations in order to maintain the target thermodynamic conditions, which can modify velocity-dependent time-correlation functions. By contrast, NVE trajectories evolve without direct thermostat or barostat coupling after equilibration and therefore provide a less perturbed representation of the intrinsic rotational dynamics. The observed ensemble dependence suggests that thermostat-induced modifications of angular-velocity correlations can lead to a systematic underestimation of SR relaxation when thermostatted trajectories are used directly for the relaxation analysis.

\begin{figure}[!htbp]
\centering
\includegraphics[width=1\linewidth]{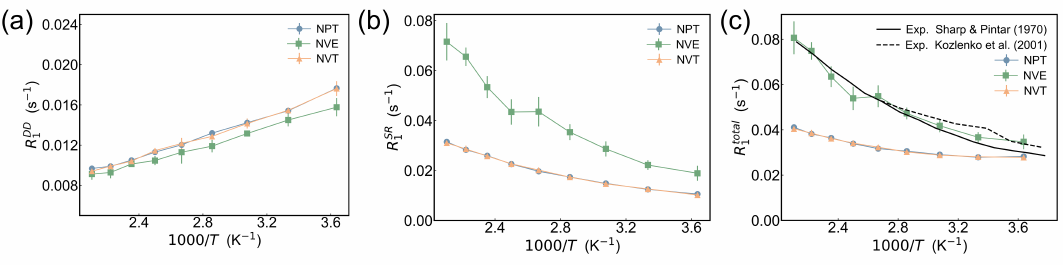}
\caption{
Comparison of the temperature-dependent $^1$H relaxation rates obtained under NpT, NVT, and NVE ensembles:
(a) dipolar relaxation,
(b) spin-rotation relaxation, and
(c) total relaxation.
}\label{figS6}
\end{figure}

\section{Vibrational and Rotational Contributions to Relaxation Based on Model~2}

\begin{figure}[!htbp]
\centering
\includegraphics[width=\linewidth]{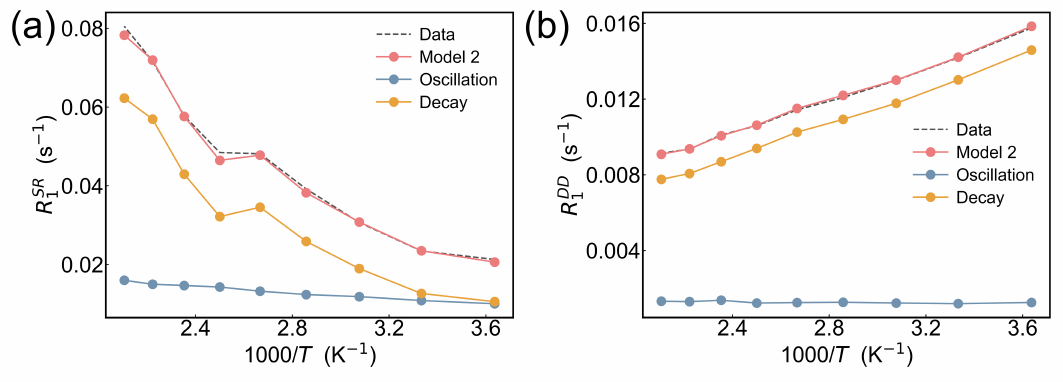}
\caption{Vibrational and rotational contributions to (a) SR and (b) DD relaxation rates determined using Model~2.}
\label{figS7}
\end{figure}

\newpage
\section{Apparent Activation Energy for NH$_4^+$ Reorientation}

\begin{figure}[!htbp]
\centering
\includegraphics[width=0.5\linewidth]{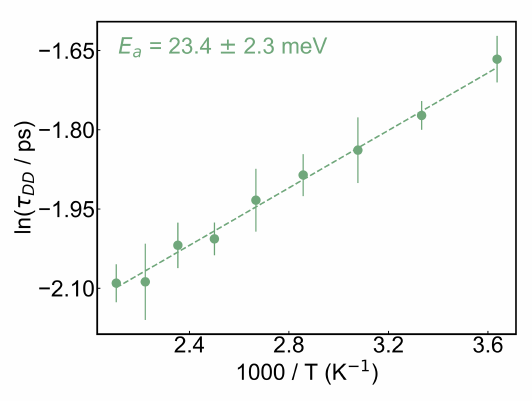}
\caption{
Arrhenius analysis of the separated rotational contribution to the dipolar correlation time.
}\label{figS8}
\end{figure}

\printbibliography